\documentclass[cmfonts,
    ,final            
 ,numberedheadings 
,bibliocites
,a4paper
  ]
  {aipproc}

\layoutstyle{6x9}

\usepackage{graphicx}
\graphicspath{pic/}
\usepackage{psfrag}

\usepackage{amssymb,amsmath} 

\usepackage[mathscr]{eucal}

\renewcommand{\Re}{\mathbb R}

\newcommand{\half}{\frac{1}{2}}

\newcommand{\la}{\langle}
\newcommand{\ra}{\rangle}

\newcommand{\TT}{\mathbb T} 
\newcommand{\calT}{\mathcal T}

\newcommand{\calM}{\mathcal M} 
\newcommand{\Riem}{\operatorname{Riem}}
\newcommand{\Ric}{\operatorname{Ric}}
\newcommand{\inj}{\operatorname{inj}}

\newcommand{\calH}{\mathcal H} 
\newcommand{\calP}{\mathcal P} 
\newcommand{\calJ}{\mathcal J} 
\newcommand{\Mbar}{\overline{M}}
\newcommand{\Kbar}{\overline{K}}
\newcommand{\Scal}{\operatorname{Sc}}
\newcommand{\Vol}{\operatorname{Vol}}

\renewcommand{\div}{\operatorname{div}}
\newcommand{\tr}{\operatorname{tr}}
\newcommand{\Trace}[1]{\smash{\sideset{^{#1}}{}{\tr}}}

\newcommand{\trSig}{\Trace{\Sigma}}

\newtheorem{theorem}{Theorem}
\newtheorem{definition}{Definition}

\newtheorem{remark}{Remark}[section]

\begin{document}

\title{The trapped region}



\author{Lars Andersson}{
  address={
Albert Einstein Institute \\
Am M\"uhlenberg 1\\
D-14467 Golm\\
Germany}
}
\renewcommand{\thefootnote}{}

\footnotetext{Based on a talk given at ERE08, Salamanca} 

\renewcommand{\thefootnote}{\arabic{footnote}}



\begin{abstract}
I will discuss 
some recent results on marginally outer trapped surfaces, 
apparent horizons 
and the trapped region. 
A couple of applications of the results developed for
marginally outer trapped surfaces 
to coalescence of black holes and to the characterization of the trapped
region are given.
\end{abstract}

\maketitle


\section{Introduction} \label{sec:intro}
Marginally outer trapped surfaces are natural candidates for quasi-local black hole boundaries in general relativity.
They are analogues of minimal surfaces in Riemannian geometry, and in
particular there is a notion of stability for marginally outer trapped
surfaces, closely related to
stability for minimal surfaces, which allows one to prove
curvature bounds analogous to those which are known for minimal surfaces. 

It is natural to consider outermost marginally outer trapped surfaces, 
which enclose every weakly outer
trapped surface. For these, we have area bounds, as
well as a replacement for the strong maximum principle, which sheds light on 
the process of black hole coalescence. 

\subsection{Notation} 
\begin{figure}[!b]
\centering
\psfrag{MAellm}{$\ell^-$}
\psfrag{MAellp}{$\ell^+$}
\psfrag{MAn}{$n$}
\psfrag{MAnu}{$\nu$}
\psfrag{MASigma}{$\Sigma$}
\centerline{\includegraphics[width=3in]{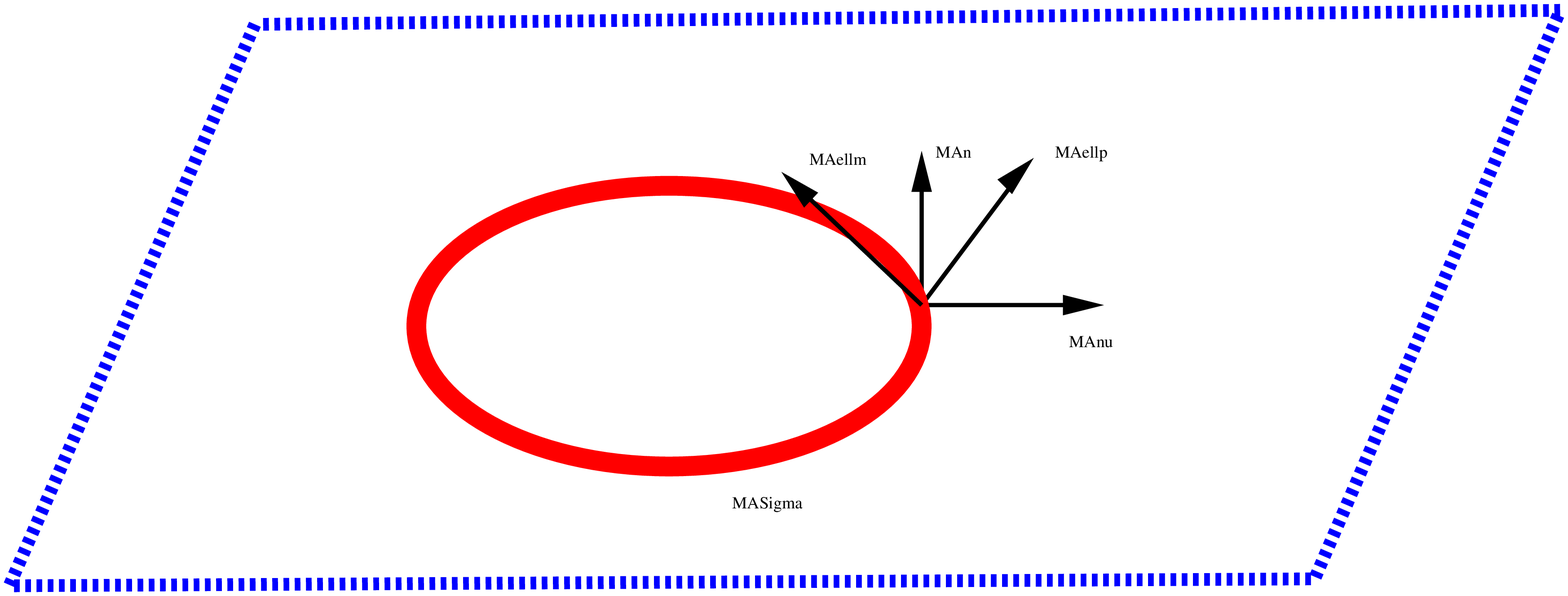}}
\caption{}
\label{fig:Sigma}
\end{figure}

Let $(M,g,K) $ be a Cauchy hypersurface in a 3+1 dimensional Lorentzian
spacetime $\calM$. Let $\Sigma$ be a spacelike surface in $M$ with null
normals $\ell^{\pm}$, see figure \ref{fig:Sigma}. 
%
We set 
\begin{align}
      A &= \la\nabla_\cdot \nu,\cdot \ra, \quad 
K^\Sigma = K|_{T\Sigma\times T\Sigma}, \nonumber \\
            \chi^\pm &= K^\Sigma \pm A,    \nonumber 
\end{align}
Then $A$ is the second fundamental form of $\Sigma$ in $M$, $K^\Sigma$ is
the restriction of $K$ to $\Sigma$, and $\chi^{\pm}$ are the null second
fundamental forms associated to $\ell^{\pm}$. Taking traces yields
$$
      H = \trSig A, \quad 
      P = \trSig K^\Sigma = \tr K - K(\nu,\nu), \quad 
      \theta^\pm  = \trSig \chi^\pm = P \pm H. 
$$
Then $H$ is the mean curvature of $\Sigma$ in $M$, $P$ is the trace of
$K^\Sigma$ on $\Sigma$ and $\theta^{\pm}$ are the {\bf null expansions} of
$\Sigma$. We declare $\ell^+$ to be the outer null normal.
\begin{definition} 
$\Sigma$ is a {\bf marginally outer trapped surface} (MOTS) if $\theta^+ = 0$.
\end{definition} 
Recall that $\theta^+$ is the logarithmic variation of area along $\ell^+$,  
$$
\theta^+ = \delta_{\ell^+} \mu_\Sigma/\mu_\Sigma
$$
If $\Sigma$ is a MOTS then 
outgoing null rays are marginally collapsing. 
We call $\Sigma$ (weakly) outer trapped if  ($\theta^+ \leq 0$) $\theta^+ <0$.
The null energy condition (NEC) holds if 
$G(v,v) \geq 0$ for any null vector $v$, where $G = \Ric - \half \Scal g$ is
the Einstein tensor. 

The usual definition of trapped surface is $\theta^+ < 0$,
  $\theta^- < 0$. By the singularity theorems of Hawking and Penrose, a
  maximal globally hyperbolic spacetime satisfying suitable energy
  conditions, eg., NEC, and which contains a trapped surface, is
  causally incomplete. We note that also the presence of an outer trapped
  surface implies incompletness. In particular, 
if NEC holds in $\calM$ and $\calM$ contains a Cauchy surface $M$ with an
outer trapped 
surface which separates and has noncompact exterior,  
then $\calM$ is null geodesically incomplete, cf. \citep{gannon,AMMS}.
Thus MOTSs may be viewed as black hole boundaries. 

The null expansion 
$\theta^+ = P + H$ is elliptic when viewed as a functional of $\Sigma \subset
M$, since, as is well known, the mean curvature $H$ has this property, and $P$ may be viewed as
a lower order term. 
For variations along $\ell^+$, 
we have 
\begin{equation}\label{eq:ray} 
\delta_{f\ell^+} \theta^+ = - Wf = - (|\chi^+|^2 + G(\ell^+,
\ell^+)) f
\end{equation} 
so $|\Sigma|$ is {\em not} an elliptic functional with respect to 
null variations. 
However, variations within $M$, of the form $\delta_{f\nu} \theta^+$ define an
elliptic operator, see section \ref{sec:stability-op} below. 

It is well known that if 
there are surfaces $\Sigma_{\pm}$ with $H[\Sigma_+]
> 0$ and $H[\Sigma_-] < 0$, which form barriers for the problem of minimizing
area, then there is a minimal ($H=0$) surface between
them. 
%
%
Suppose we have an analogue for MOTSs of existence in the presence of
barriers, in this case surfaces $\Sigma_{\pm}$ with the inner barrier
satisfying $\theta^+[\Sigma_-] < 0$,
while the outer barrier satisfies $\theta^+[\Sigma_+] > 0$. 
Then, in view of 
(\ref{eq:ray}), MOTSs should {\em persist} if NEC holds. This can in fact be
proved using the results of \citep{Andersson:2007gy}, given an outer barrier,
cf. \citep{AMMS}, see section \ref{sec:MOTSpersist}. 
In particular, if $(M,g,K)$ is an asymptotically flat initial data set, then
there is an outer barrier in $M$. We shall, throughout the rest of this
note, assume the presence of an outer barrier.

Due to the persistence of MOTSs, we expect that MOTSs are generically in a {\bf 
marginally outer trapped tube} (MOTT), i.e. a hypersurface of $\calM$, 
foliated
by MOTSs. This has been proved for
outermost MOTSs, modulo a genericity condition, cf. theorem 
\ref{thm:AMS05AMMS} and \citep{AMMS}. In this case, the MOTT is weakly
spacelike if NEC holds. 
If in addition $\theta^- < 0$
on the MOTS, the MOTT is a dynamical horizon, cf. 
\citep{ashtekar:krishnan,ashtekar:galloway}. 


\subsection{The stability operator}\label{sec:stability-op} 
Let $Lf = \delta_{f \nu} \theta^+$. 
Then 
$$
Lf = - \Delta f + 2 S(\nabla f) + f [ \div S - |S|^2 - \half | \chi^+|^2 +
  \half {}^{\Sigma} \Scal + (\mu - J(\nu))]
$$
where $S(X) = K(X,\nu)$. 
The operator $L$ is the analogue of the minimal surface stability operator. 
We note the following facts which hold for $L$. 
The operator $L$ is 2:nd order elliptic and non-self adjoint in
general. There is a unique principal eigenvalue $\lambda \in \Re$, 
with positive eigenfunction $\phi$. If $\Sigma$ is a locally outermost MOTS then
$\lambda \geq 0$.
Further, 
if $\lambda \geq 0$ then there is $f \geq 0$ such that $Lf \geq 0$, i.e., 
if $\lambda \geq 0$ the maximum principle holds. 

\begin{definition} $\Sigma$ is {\bf stable} if $\lambda \geq 0$. 
\end{definition} 
In particular, if
$\Sigma$ is locally outermost then $\Sigma$ is stable. 

\subsection{Local existence of horizons}

The proof of the following theorem makes use of the definition of $L$, the 
implicit function theorem, as well as the above mentioned version of the
maximum principle. 
The following theorem was proved in \citep{Andersson:2005gq,AMMS}. 
\begin{theorem}
\label{thm:AMS05AMMS}
Suppose $\Sigma$ is stable
  ($\lambda \geq 0$). If $\lambda = 0$, assume in addition\footnote{The
  condition that 
$W$ is not identically zero 
may be viewed as a genericity condition} 
$W$ is not identically zero.
Then $\exists$ a MOTT $\mathbf H$ containing $\Sigma$. $\mathbf H$ is
  weakly spacelike if NEC holds. 
\end{theorem} 
\begin{figure}[!bpt]
\centering
\psfrag{MAthetal0}{$\theta^+ \leq 0$}
\psfrag{MAthetag0}{$\theta^+ \geq 0$}
\psfrag{MAMOTTH}{MOTT $\mathbf H$}
\psfrag{MAJump}{outermost MOTS after jump}
\psfrag{MAellp}{$\ell^+$} 
\psfrag{MAMt}{$M_t$}
\centerline{\includegraphics[width=3in]{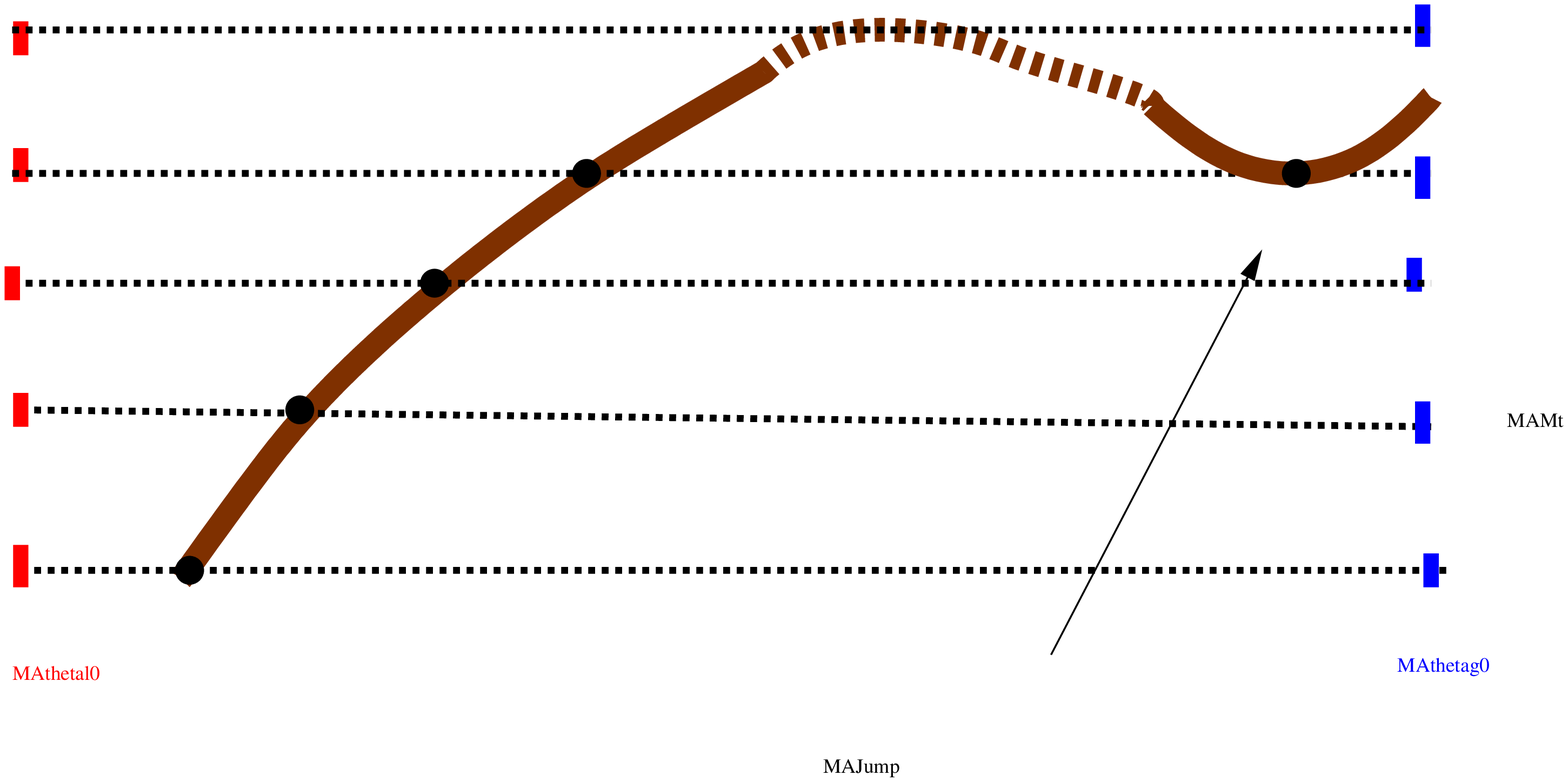}}
\caption{}
\label{fig:mott3}
\end{figure}
Theorem \ref{thm:AMS05AMMS} is a local result. In general, 
the outermost MOTS can jump. This may happen for example through the
formation of a new MOTS outside the existing ones, a process that can be
caused by the coalescence of black holes, cf. section \ref{sec:coalescence}. 
As a MOTS is created, the MOTT bifurcates in general, see
\citep{AMMS}, see also figure \ref{fig:mott3}. 

\subsection{Curvature bounds}
\begin{theorem}[\citep{Andersson:2005me}] \label{thm:AM05} 
Let $\Sigma$ be a stable MOTS. Then
$$
|A| \leq C(|\Riem|_{C^0}, |K|_{C^1}, \inj(M))
$$
\end{theorem} 
The proof of theorem \ref{thm:AM05} 
applies several techniques used  in  \citep{SSY}, including   
the Simons identity, a Kato inequality, and the Hoffmann-Spruck 
Sobolev inequality. The Moser iteration used in \citep{SSY} to achieve the
$L^\infty$ estimate for $|A|$ is replaced by a Stampacchia
iteration. Further, the symmetrized stability estimate of 
\citep{galloway:schoen} is used. By applying the local area bound of
Pogorelov, it is possible to avoid a dependence on the area 
$|\Sigma|$ in the curvature estimate. Due to the use of this result, which in
turn relies upon the Gauss-Bonnet theorem, the above form of the curvature
estimate applies only to the case of a 2-dimensional surface in a 3+1
dimensional spacetime. 

\section{Existence of MOTSs}

\subsection{Jang's equation}
\begin{figure}[!bpt]
\centering
\psfrag{MAs}{$s$}
\psfrag{MAsf}{$\Mbar = \{s=f\}$}
\psfrag{MAnu}{$\nu$}
\psfrag{MAMb}{}
\psfrag{MAM}{$M$}
\centerline{\includegraphics[width=2in]{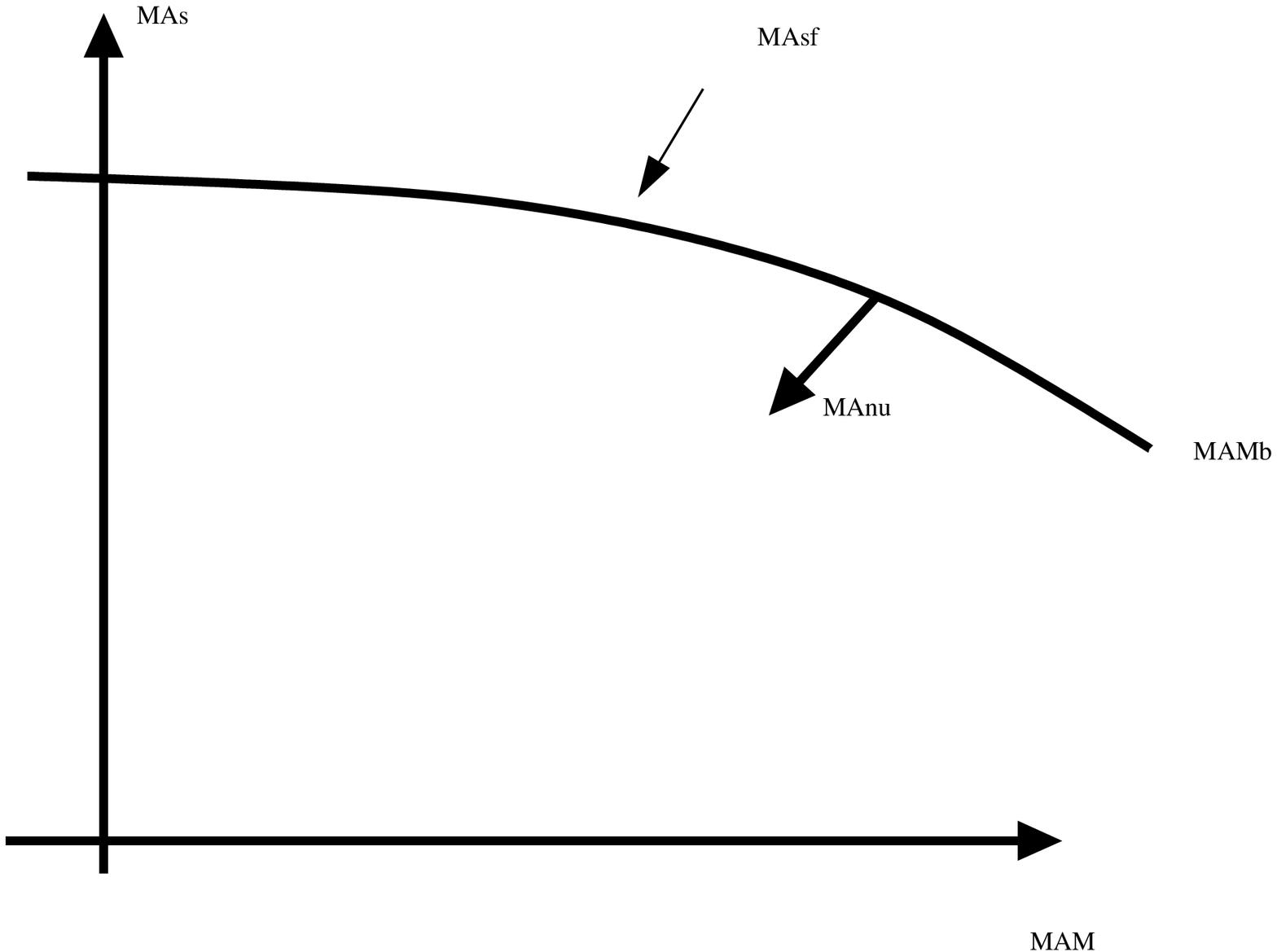}}
\caption{}
\label{fig:jang}
\end{figure}

Consider $\Re \times M$, 
with metric $ds^2 + g$, see figure \ref{fig:jang}. 
Define $\Kbar$ by pullback of $K$.
Let $\Mbar$ be the graph of $f$. 
On $\Mbar$ we have induced mean curvature $\calH$ and 
$\calP = \tr_{\Mbar} \Kbar$. 
Jang's equation is 
$$
\calJ [f] := \calH - \calP = 0 
$$
This is the analogue of the equation $\theta^+ = 0$

\begin{figure}[!bpt]
\centering 
\begin{minipage}{0.45\textwidth}
\centering
\psfrag{MAnu}{$\nu$}
\psfrag{MAM}{$M$}
\psfrag{MAthp0}{$\theta^+=0$}
\includegraphics[width=2.2in]{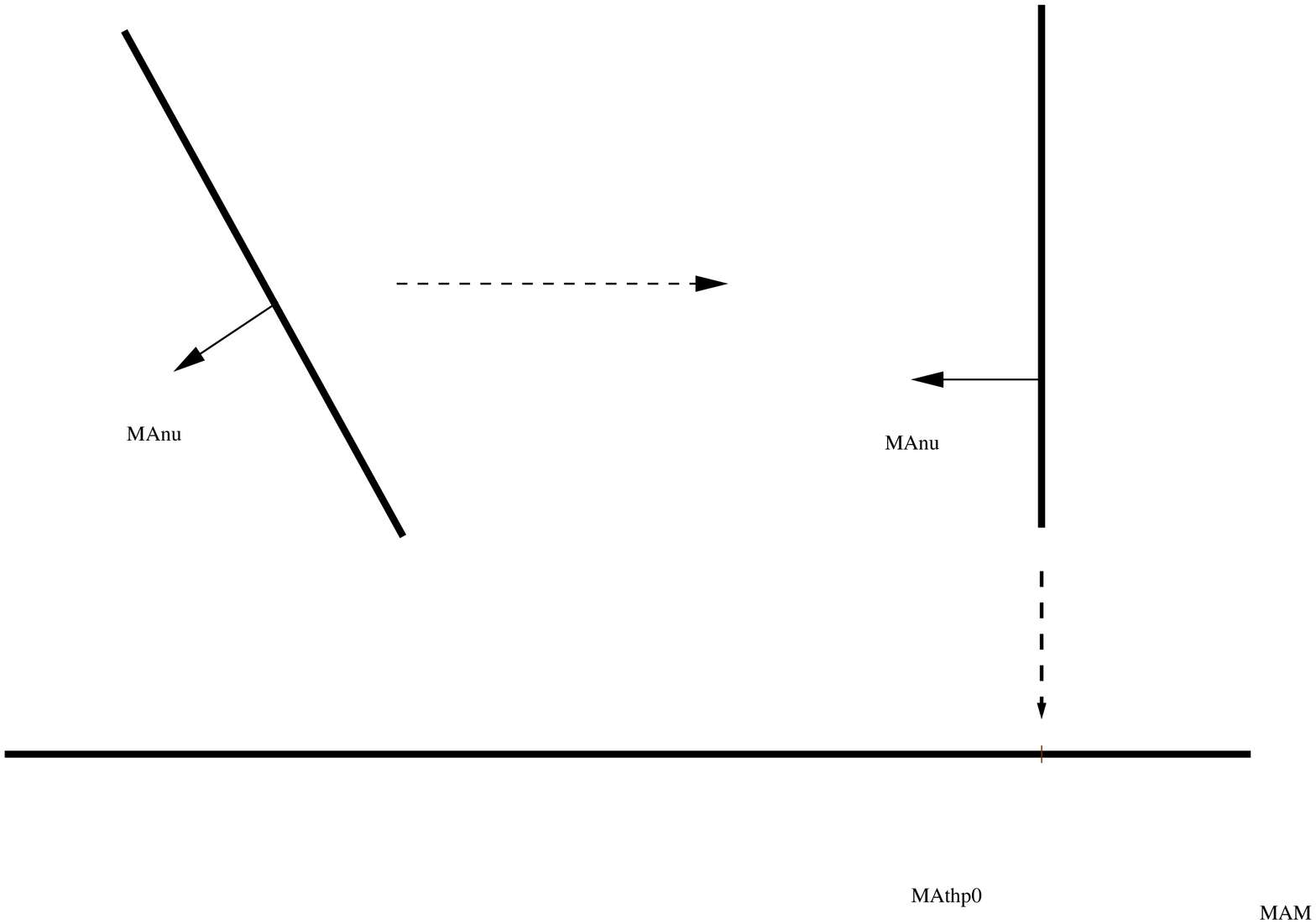}
\end{minipage}%
\begin{minipage}{0.45\textwidth}
\centering 
\psfrag{MAnu}{$\nu$}
\psfrag{MAM}{$M$}
\psfrag{MAthm0}{$\theta^-=0$}
\includegraphics[width=2.2in]{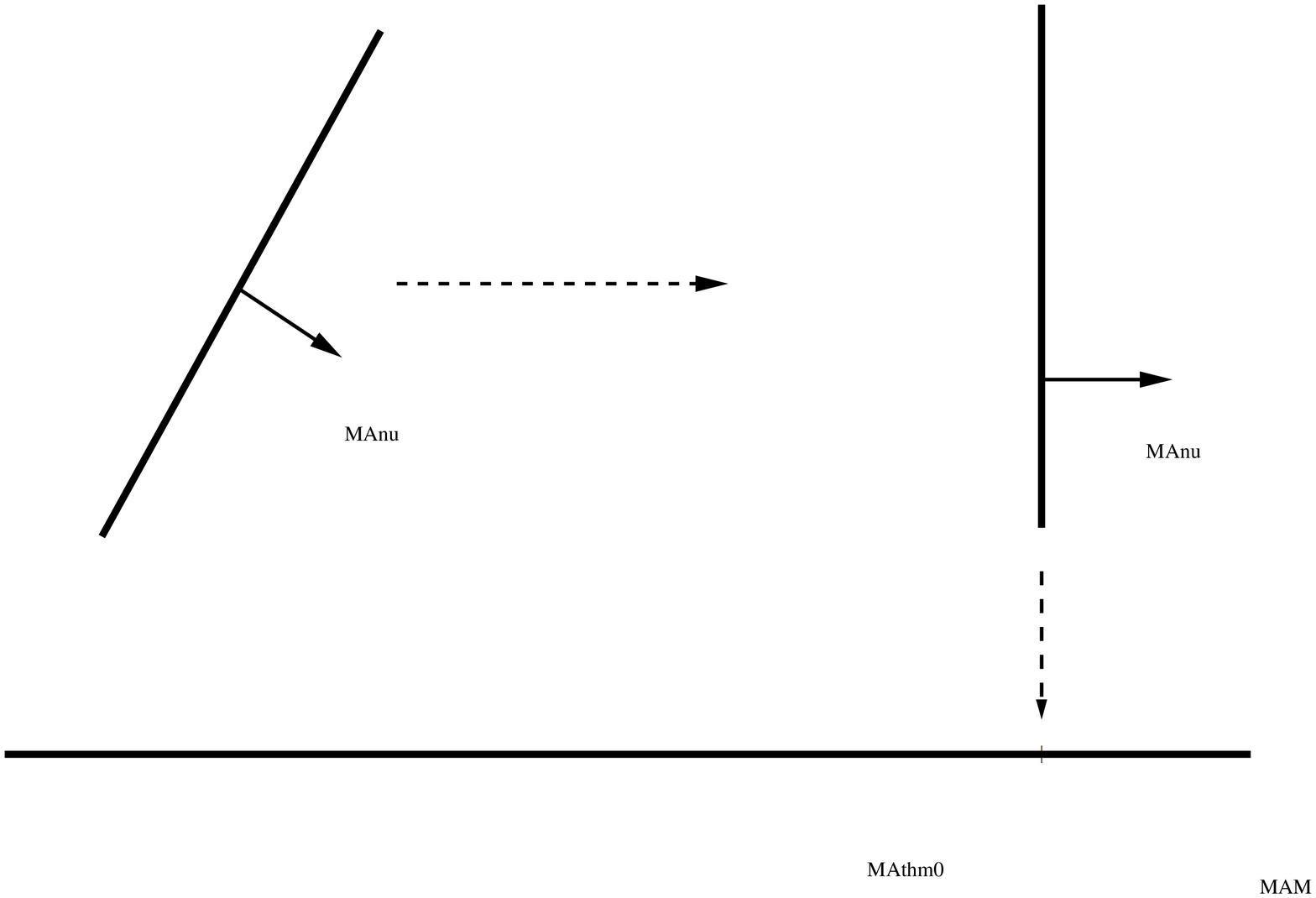}
\end{minipage}
\caption{}
\label{fig:jangblowMOTS}
\end{figure}

\begin{figure}[!bpt]
\centering
\psfrag{MAZ}{$Z$}
\psfrag{MAmZ}{$-Z$}
\psfrag{MAdmM}{$\partial^- M$}
\psfrag{MAdpM}{$\partial^+ M$}
\psfrag{MAcalJ0}{$\calJ = 0$}
\includegraphics[width=2in]{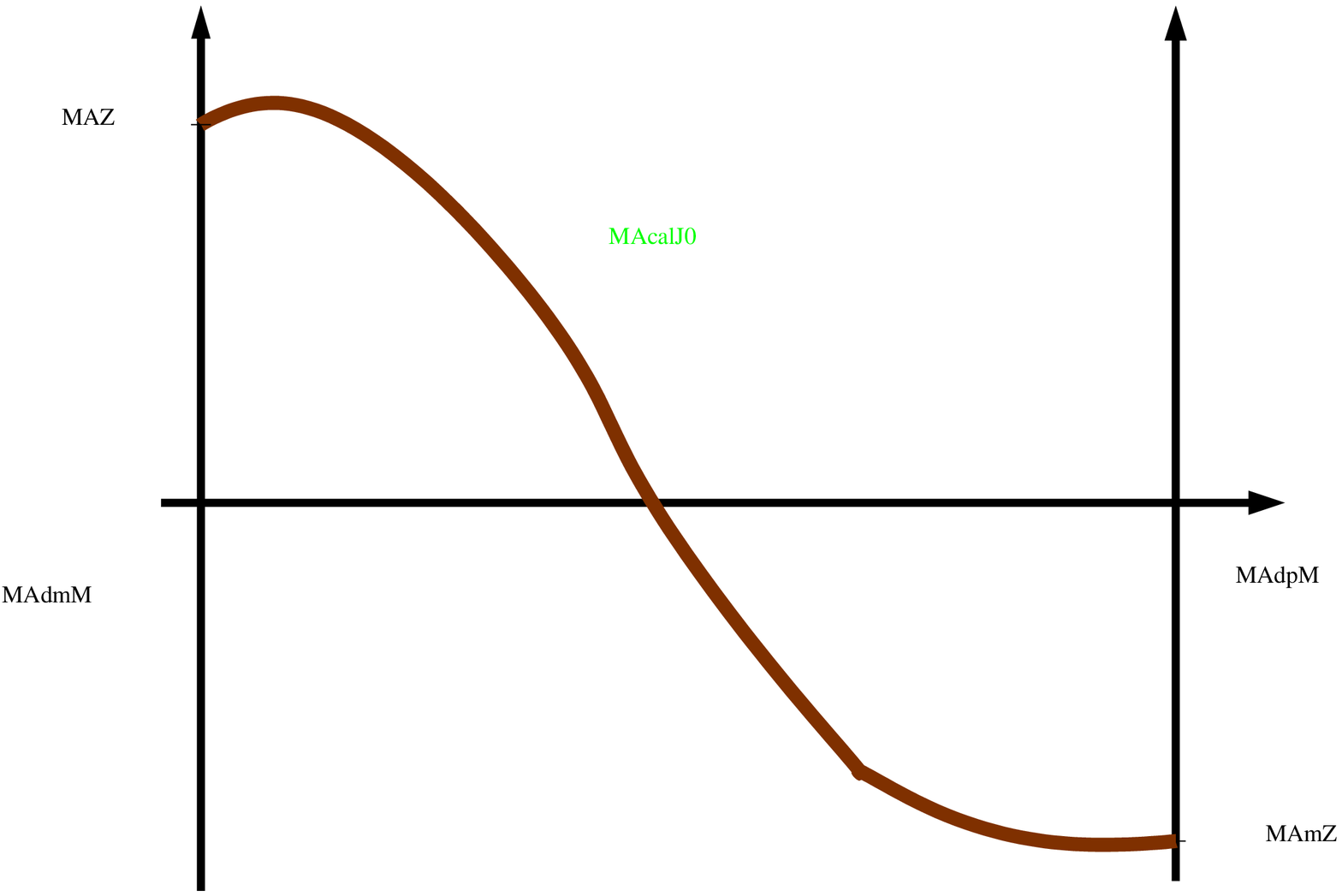}
\caption{}
\label{fig:jangdirichlet}
\end{figure}

By translation invariance,  $J[f] = J[f + t]$, we have that the 
stability operator for $\Mbar$ has $L \phi = 0$, 
with $\phi = \la \nu , e_4 \ra$. Here
$L$ is the analogue of the minimal surface stability operator for $\Mbar$.
By the work in \citep{SYII}, we have 
local curvature bounds for $\Mbar$, which yields compactness. This allows one
to prove existence of solutions to Jang's equation using 
a capillarity deformation together with Leray-Schauder theory. One considers
the deformed equation 
$$
\calH - \sigma \calP = \tau f_{\sigma,\tau}
$$
As 
$\sigma$ goes from 0 to 1, we have a limit $f_\tau$. Letting 
$\tau \searrow 0$, we have by compactness, convergence of a 
subsequence of $f_\tau$ to a solution $f$ of Jang's equation. 
The solution has blowups in general. As observed in \citep{SYII}, 
blowups project to MOTSs, cf. figure \ref{fig:jangblowMOTS}.
Therefore Jang's equation can be used to prove existence of MOTSs. It was
proved in \citep{Andersson:2007gy} that the blowup surfaces 
are {\em stable} MOTSs.
\begin{theorem}[\citep{RS,Andersson:2007gy}] \label{thm:MOTSexist} 
Suppose $M$ is compact with
  barrier boundaries $\partial^{\pm} M$, such that 
$$
\theta^+[\partial^- M] < 0, \quad \theta^+[\partial^+ M] > 0
$$
Then $M$ contains a MOTS $\Sigma$.
\end{theorem} 
Theorem \ref{thm:MOTSexist} provides 
the analogue of the barrier argument for existence of minimal
surfaces.
The proof considers a sequence of Dirichlet problems for Jang's
equation, which forces a blowup solution. 
We solve 
$$
\calJ[f] = 0, \quad 
f \bigg{|}_{\partial^{\pm} M} = \mp Z
$$
see figure \ref{fig:jangdirichlet}. 
If we let $Z \to \infty$, then the solution converges to 
solution with blowups. 

\begin{figure}[!bpt]
\centering 
\psfrag{MAtM}{$\tilde M$}
\psfrag{MAdpM}{$\partial^+ M$}
\psfrag{MAthpg0}{$\theta^+ > 0$}
\psfrag{MAMgK}{$(M,g,K)$}
\psfrag{MAHg0}{$H > 0$}
\includegraphics[width=2in]{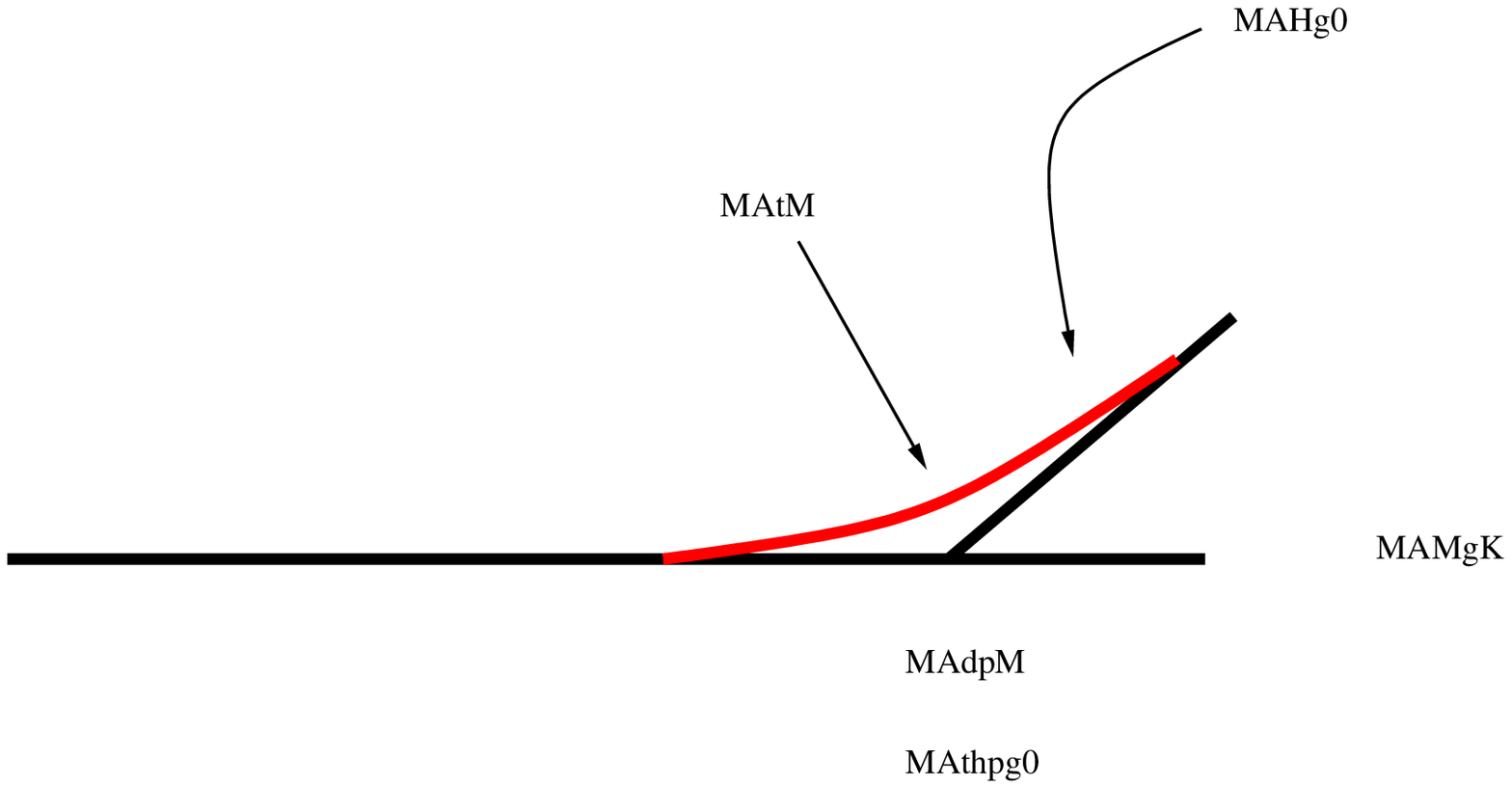}
\caption{}
\label{fig:jangbend} 
\end{figure}

In order to prove boundary gradient estimates necessary to apply
Leray-Schauder theory, one makes use of a deformation of the Cauchy surface,
see figure \ref{fig:jangbend}, 
to get $H> 0$ at $\partial M$.
The limiting solution must blow up somewhere, which implies the existence of
a MOTS. We have a foliation by barriers near $\partial M$. Using this fact,
and the maximum principle, one can show the MOTSs constructed are in 
the {\em undeformed} region of $M$.  

By deforming the data inside $\partial^- M$, we can 
allow $\theta^+[\partial^- M] \leq 0$. 
We remark that \citep{eichmair-2007} has studied the Plateau problem for MOTSs using Perron's
method. 

\subsection{Application: Persistence of MOTSs} \label{sec:MOTSpersist} 
By Rauchaudhuri, $\delta_{\ell^+} \theta^+ = - W \leq 0$ if NEC holds. 
\begin{theorem}[\citep{AMMS}] \label{thm:AMMS}
Let $\calM$ be a spacetime which satisfies
  NEC. Let $\{M_t\}$,  
be a Cauchy foliation of $\calM$, and 
  assume we have outer barriers. 

If $M_0$ contains a MOTS, then each $M_t$, $t \geq 0$ contains a MOTS. 
\end{theorem} 
Based on theorem \ref{thm:AMMS},  
it seems natural to view the collection of outermost 
MOTS in $M_t$ as the black hole boundary in
$\calM$. If it is smooth, this collection is a MOTT. 
For further regularity and continuation results for MOTTs, cf. \citep{AMMS}.

\section{Area bound}

\begin{theorem}[\citep{Andersson:2007gy}] \label{thm:areabound} 
Suppose $M$ has an outer barrier. There is a constant $C = C(|\Riem|_{C^0},
|K|_{C^1}, \inj(M), \Vol(M))$ such that for a bounding 
MOTS $\Sigma$ in $M$, either 
$$
|\Sigma| \leq C
$$
or 
there is a MOTS $\Sigma'$ {\em outside} $\Sigma$. 
\end{theorem} 
The idea of proof of theorem \ref{thm:areabound} is the following. 
If $|\Sigma|$ is very large, then due to curvature bounds and the bounded
$\Vol(M)$, $\Sigma$ must nearly meet itself from the outside, which implies
that the outer
injectivity radius $i^+(\Sigma)$ must be small. In this situation we can
use surgery and heat flow to show the existence of a MOTS outside $\Sigma$. 
\begin{figure}[!bpt]
\centering
\begin{minipage}{0.55\textwidth}
\centering
\psfrag{MASigma}{$\Sigma$} 
\psfrag{MAnu}{$\nu$}
\includegraphics[width=\textwidth]{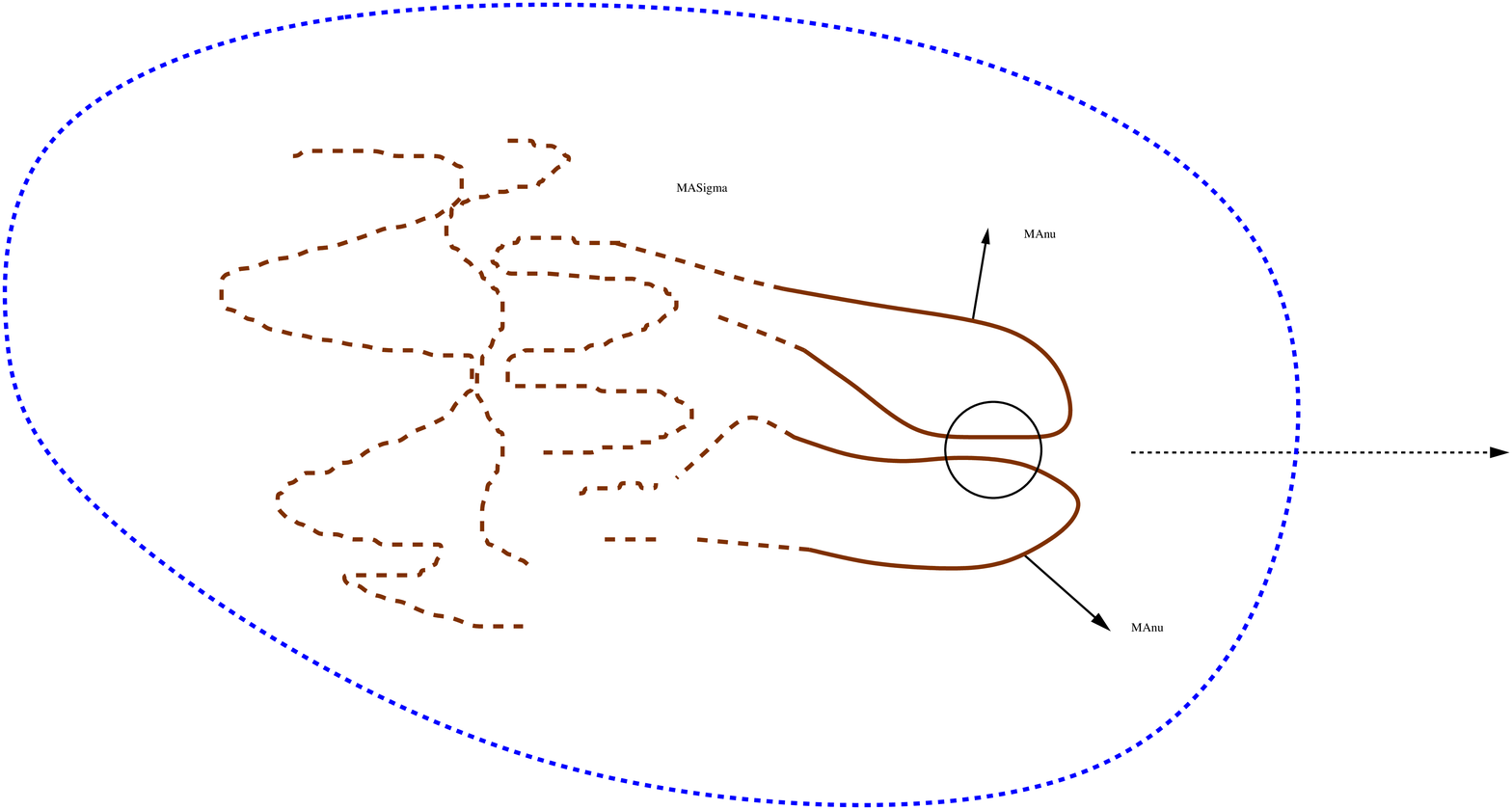}
\end{minipage}%
\begin{minipage}{0.45\textwidth}
\centering
\psfrag{MAthpl0}{$\theta^+ < 0$}
\psfrag{MAnu}{$\nu$}
\psfrag{MAipSigma}{$i_+(\Sigma)$}
\includegraphics[width=.9\textwidth]{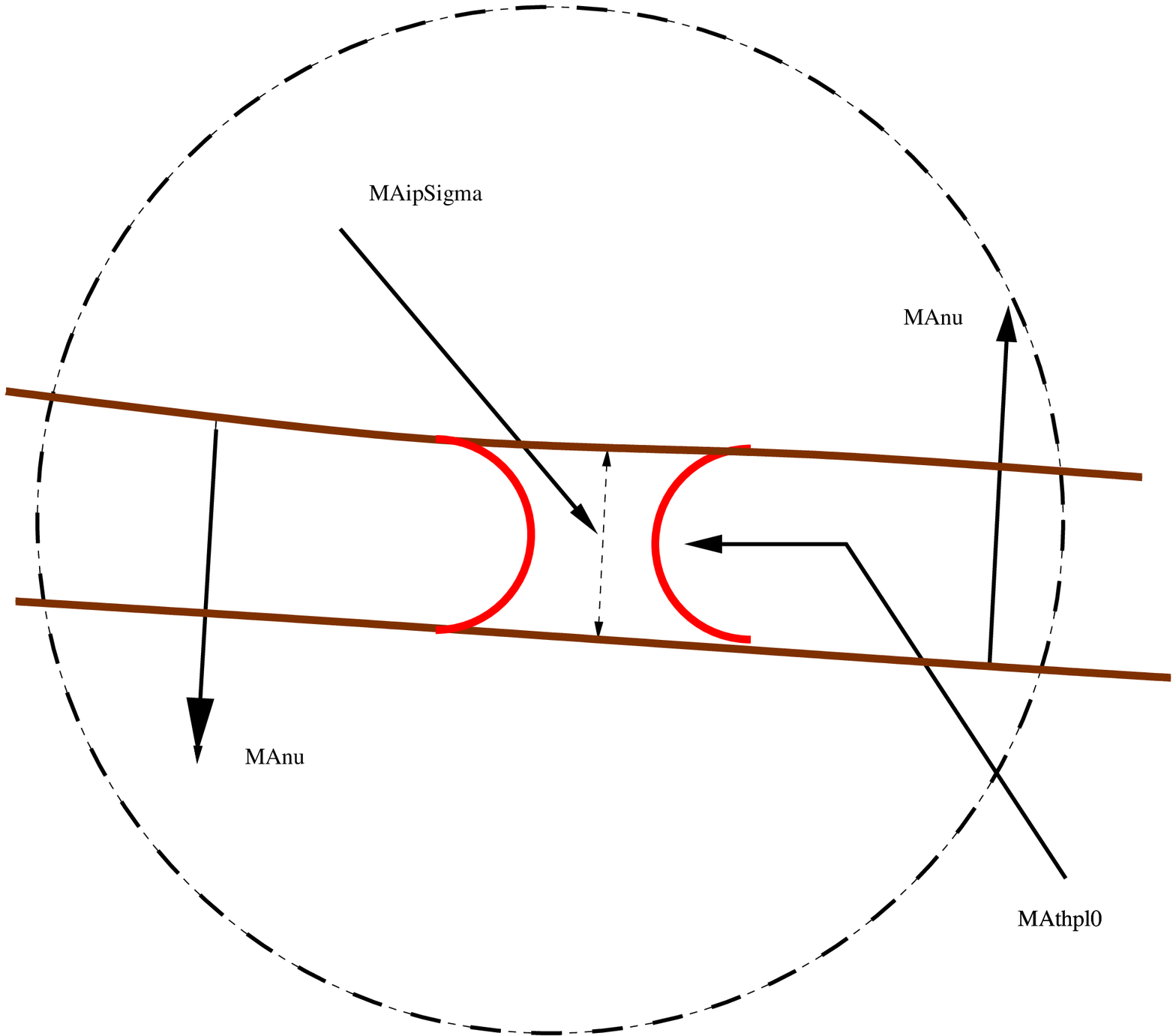}
\end{minipage} 
\caption{Gluing in a neck with $\theta^+ < 0$.} 
\label{fig:volumefigs} 
\end{figure}
In a location where $\Sigma$ nearly meets itself on the outside, we 
glue in a neck with $\theta^+ < 0$, see figure \ref{fig:volumefigs}. 
The resulting surface is then deformed
using the $\theta^+$ heat flow 
$$
\dot x = - \theta^+ \nu,
$$ 
this gives a family $\Sigma_s$, $s \geq
0$. 
The maximum
principle can be used to show that for $s > 0$, 
$\Sigma_s $ is outside $\Sigma$, 
with $\theta^+[\Sigma_s] < 0$. 
Thus, 
$\Sigma_s$ is an inner barrier, which means that we can apply the 
existence result theorem \ref{thm:MOTSexist}. It follows there is a 
MOTS $\Sigma_{\text{new}}$ {\em outside} $\Sigma$.
Each time the above argument is applied it uses at least 
$\delta_{\Vol}$ of the volume outside $\Sigma$,
%
and hence after finitely many steps, one 
has a $\Sigma_0$ outside $\Sigma$ with outer
  injectivity radius $i^+(\Sigma_0) > \delta_*$. 
The surface 
$\Sigma_0$ has the claimed area bound. To estimate the area, we use the
  estimates on curvature and $i^+$ to estimate the volume of a tube around
  $\Sigma_0$ from below, using the divergence theorem, 
in terms of $|\Sigma_0|$. 
This tube must have volume bounded by $\Vol(M)$, which leads to an estimate
for $|\Sigma_0|$.  This area bound, together with the curvature bound from
Theorem \ref{thm:AM05}, gives compactness for the family of outermost MOTSs 
in a sequence of Cauchy data sets with suitable uniformity properties,
cf. \citep{Andersson:2007gy}.  

\subsection{Application: Coalescence of black holes} \label{sec:coalescence} 


It is a direct consequence of the gluing and heat flow construction used in
the proof of the area bound, that if $\Sigma_1, \Sigma_2$ are locally
outermost MOTSs which are sufficiently close, then there
is a MOTS $\Sigma$ surrounding them. This may be interpreted as stating that
black holes must coalesce once they are sufficiently close.
The phenomenon
described here is seen in numerical simulations. As the MOTS $\Sigma$ is
formed in an evolution, it has principal eigenvalue $\lambda = 0$ and
generically there is a MOTT which bifurcates into existence when $\Sigma$ is
formed. See \citep{AMMS} for details. 
\begin{remark} The above result gives a {\em ``maximum principle for MOTS''}. 
Note that the usual
maximum principle {\em does not apply} for MOTSs which meet on the
{\em outside}. 
\end{remark} 

\section{The trapped region}

\begin{definition}Let $(M,g,K)$ be an AF data set. The trapped region is 
$$
\TT = \cup_{\Omega \subset M} \{ \partial \Omega \quad \text{ is weakly outer
  trapped} \} 
$$
\end{definition} 

\begin{theorem}[\citep{Andersson:2007gy}] If $\exists$ $\Omega \subset M$,
  with $\partial \Omega$ weakly outer trapped, then $\TT$ has smooth boundary 
$\partial \TT$, with $\theta^+[\partial \TT] = 0$. In particular, $\partial
  \TT$ is the {\em unique outermost MOTS in $M$}.
\end{theorem} 
For the proof, replace $\TT$ by 
$$
\calT = \cup_{\Omega} \{ \theta^+[\partial \Omega] \leq 0, \quad \text{ and } 
i^+(\partial \Omega) \geq \delta_* \} 
$$
For the collection of subsets defining $\calT$ we have {\em compactness} by
the area bound for surfaces with $i^+$ bounded from below. This, together
with a gluing construction to smooth corners gives that 
$\partial \calT = \Sigma$ is a 
MOTS.  To complete the proof we have to show that $\TT \subset \calT$. 
To see this, suppose there is a weakly outer trapped surface 
$\Omega \nsubseteq \calT$. We can argue
that this means $\Omega \cap \calT \ne 0$. 
\begin{figure}[!bpt]
\centering
\psfrag{MAsmooth}{$\theta^+ < 0$ by smoothing} 
\psfrag{MAOmega}{$\Omega$}
\psfrag{MAcalT}{$\calT$}
\centerline{\includegraphics[width=2.5in]{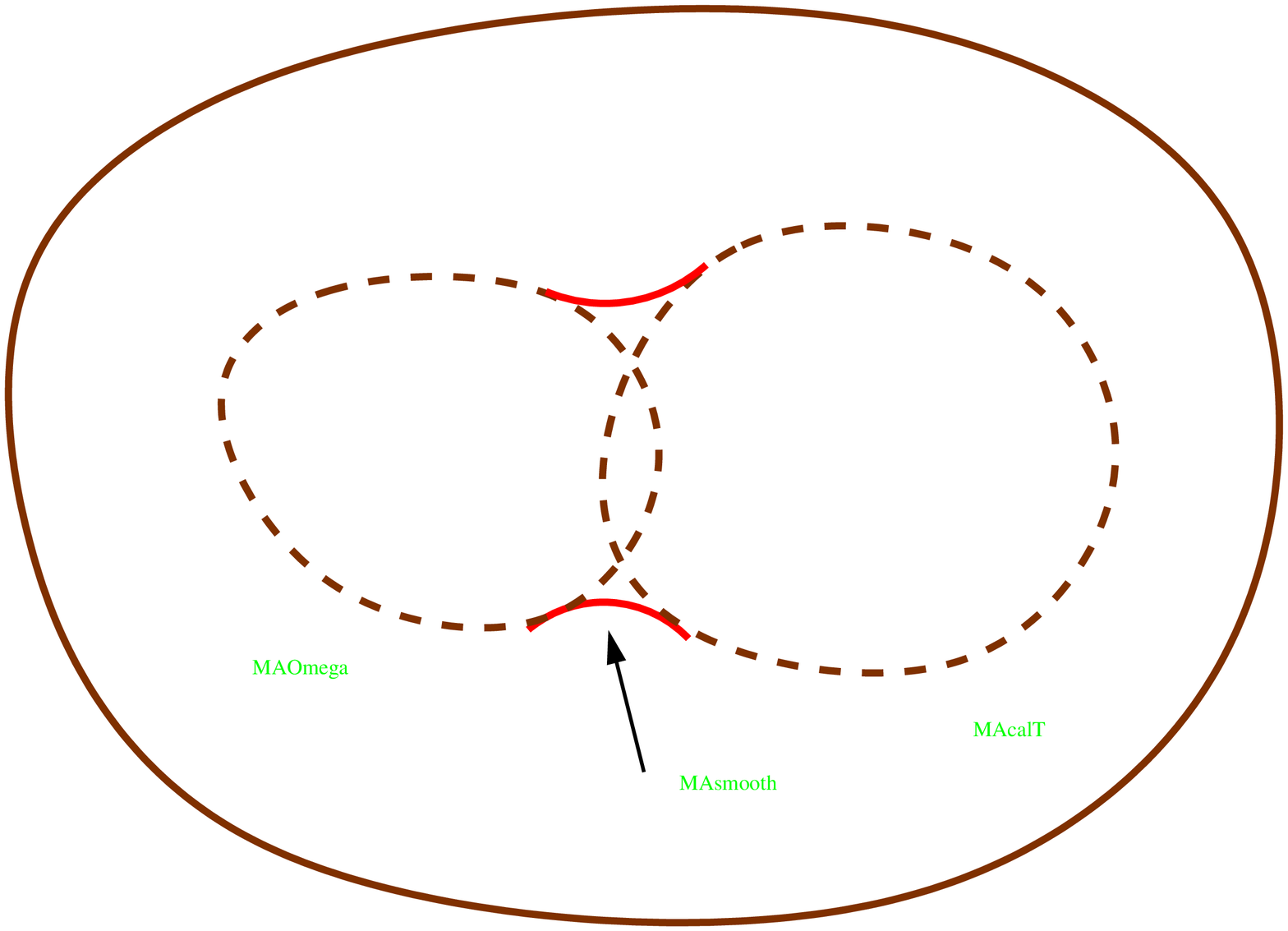}}
\caption{}
\label{fig:smoothing2} 
\end{figure}
Smoothing gives a barrier, see figure \ref{fig:smoothing2}  
and hence there is a MOTS outside. This can be
taken to be in $\calT$. Hence $\TT = \calT$, which completes the proof. 

\begin{remark}
By \citep{galloway:2007}, the 
outermost MOTS is a union of finitely many $S^2$, assuming NEC. 
\end{remark}

\section{Concluding remarks}
Bray and Khuri 
have proposed generalized apparent horizons (GAH), satisfying the condition 
$H = |P|$ as well as a 
generalized Jang's equation motivated by the GAH condition, 
as part of an approach to the general Penrose
Inequality. Eichmair \citep{eichmair-2008} proved
  existence of outermost GAH.
These are area outer minimizing. However, it is not clear
  how they are related to black holes. 
Large families of GAH conditions can be treated using the techniques
  discussed here. 

Global properties of MOTTs may be relevant for understanding the strong
  field Cauchy problem for the Einstein equations. 

The known conditions for existence of MOTSs in a Cauchy data set 
\citep{SY:condensation,yau:existBH,Galloway:2008gc} involve nonvacuum data. 
A better understanding of conditions for the existence of MOTSs in
  vacuum, due to concentration of curvature in 
  terms of, say, curvature radii, conformal spectral gap, etc. is needed. 

\subsection*{Acknowledgements}
%
I thank the Mittag-Leffler-Institute, Djursholm, Sweden
for hospitality and support. 
This work was supported in part by the NSF, under
contract no.\ DMS 0407732 and DMS 0707306 with the University of
Miami. 


\begin{thebibliography}{17}
\expandafter\ifx\csname natexlab\endcsname\relax\def\natexlab#1{#1}\fi
\providecommand{\enquote}[1]{``#1''}
\expandafter\ifx\csname url\endcsname\relax
  \def\url#1{\texttt{#1}}\fi
\expandafter\ifx\csname urlprefix\endcsname\relax\def\urlprefix{URL }\fi
\providecommand{\eprint}[2][]{\url{#2}}

\bibitem[Gannon(1976)]{gannon}
D.~Gannon, \emph{General Relativity and Gravitation} \textbf{7}, 219--232
  (1976).

\bibitem[Andersson et~al.(2008)]{AMMS}
L.~Andersson, M.~Mars, J.~Metzger, and W.~Simon  (2008), \eprint{0811.4721}.

\bibitem[Andersson and Metzger(2007)]{Andersson:2007gy}
L.~Andersson, and J.~Metzger  (2007), \eprint{0708.4252}.

\bibitem[Ashtekar and Krishnan(2002)]{ashtekar:krishnan}
A.~Ashtekar, and B.~Krishnan, \emph{Phys. Rev. Lett.} \textbf{89}, 261101, 4
  (2002).

\bibitem[Ashtekar and Galloway(2005)]{ashtekar:galloway}
A.~Ashtekar, and G.~J. Galloway, \emph{Adv. Theor. Math. Phys.} \textbf{9},
  1--30 (2005).

\bibitem[Andersson et~al.(2005)]{Andersson:2005gq}
L.~Andersson, M.~Mars, and W.~Simon, \emph{Phys. Rev. Lett.} \textbf{95},
  111102 (2005), \eprint{gr-qc/0506013}.

\bibitem[Andersson and Metzger(2005)]{Andersson:2005me}
L.~Andersson, and J.~Metzger  (2005), \eprint{gr-qc/0512106}.

\bibitem[Schoen et~al.(1975)]{SSY}
R.~Schoen, L.~Simon, and S.~T. Yau, \emph{Acta Math.} \textbf{134}, 275--288
  (1975).

\bibitem[Galloway and Schoen(2006)]{galloway:schoen}
G.~J. Galloway, and R.~Schoen, \emph{Comm. Math. Phys.} \textbf{266}, 571--576
  (2006).

\bibitem[Schoen and Yau(1981)]{SYII}
R.~Schoen, and S.~T. Yau, \emph{Comm. Math. Phys.} \textbf{79}, 231--260
  (1981).

\bibitem[Schoen(2004)]{RS}
R.~Schoen  (2004).

\bibitem[Eichmair(2007)]{eichmair-2007}
M.~Eichmair, The {Plateau} problem for apparent horizons (2007),
  arXiv.org:0711.4139.

\bibitem[Galloway(2008)]{galloway:2007}
G.~J. Galloway, \emph{Comm. Anal. Geom.} \textbf{16}, 217--229 (2008).

\bibitem[Eichmair(2008)]{eichmair-2008}
M.~Eichmair, Existence, regularity, and properties of generalized apparent
  horizons (2008), arXiv.org:0805.4454.

\bibitem[Schoen and Yau(1983)]{SY:condensation}
R.~Schoen, and S.~T. Yau, \emph{Comm. Math. Phys.} \textbf{90}, 575--579
  (1983).

\bibitem[Yau(2001)]{yau:existBH}
S.~T. Yau, \emph{Adv. Theor. Math. Phys.} \textbf{5}, 755--767 (2001).

\bibitem[Galloway and O'Murchadha(2008)]{Galloway:2008gc}
G.~J. Galloway, and N.~O'Murchadha, \emph{Class. Quant. Grav.} \textbf{25},
  105009 (2008), \eprint{0802.3247}.

\end{thebibliography}

\end{document}